\documentclass[onecolumn, amsmath,amssymb, aps]{revtex4-2}

\usepackage{graphicx}
\usepackage{dcolumn}
\usepackage{bm}

\usepackage{amsmath}
\usepackage{amscd}
\usepackage{amssymb}
\usepackage{amsfonts}
\usepackage[mathscr]{euscript}
\usepackage{pstricks}
\usepackage{pst-node}
\usepackage{epsfig}
\usepackage{subcaption}
\usepackage{braket}
\usepackage{tikz}
\usetikzlibrary{quantikz}
\usepackage[T1]{fontenc}

\begin{document}

\preprint{Quantum Science \& Tech.  2024 - RQE}

\title{Characterization and thermometry of dissipatively stabilized steady states}

\author{G. S. Grattan}%
\author{A. M. Liguori-Schremp}
 \email{aliguori@mines.edu}
\author{D. Rodriguez Perez}
\author{E. Kapit}
\affiliation{%
 Physics Department, Colorado School of Mines \\
 1523 Illinois Street \\ 
 Golden, CO 80401
}%

\author{W. Jones}%
\author{P. Graf}%

\affiliation{
 National Renewable Energy Laboratory \\ 
 15013 Denver W Pkwy \\ 
 Golden, CO 80401
}%

\date{\today}

\begin{abstract}
In this work we study the properties of dissipatively stabilized steady states of noisy quantum algorithms, exploring the extent to which they can be well approximated as thermal distributions, and proposing methods to extract the effective temperature T. We study an algorithm called the Relaxational Quantum Eigensolver (RQE), which is one of a family of algorithms that attempt to find ground states and balance error in noisy quantum devices. In RQE, we weakly couple a second register of auxiliary “shadow” qubits to the primary system in Trotterized evolution, thus engineering an approximate zero-temperature bath by periodically resetting the auxiliary qubits during the algorithm’s runtime. Balancing the infinite temperature bath of random gate error, RQE returns states with an average energy equal to a constant fraction of the ground state. We probe the steady states of this algorithm for a range of base error rates, using several methods for estimating both T and deviations from thermal behavior. In particular, we both confirm that the steady states of these systems are often well-approximated by thermal distributions, and show that the same resources used for cooling can be adopted for thermometry, yielding a fairly reliable measure of the temperature. These methods could be readily implemented in near-term quantum hardware, and for stabilizing and probing Hamiltonians where simulating approximate thermal states is hard for classical computers.
\end{abstract}

\maketitle

\section{Introduction}

Many important applications of quantum computation involve preparing and/or approximating the ground state of quantum Hamiltonians, for which many algorithms have been proposed, ranging from variational methods such as the Variational Quantum Eingensolver (VQE) \cite{dallaire-demers-2020, tilly-PRep-2022} to techniques based on adiabatic processes \cite{farhi-sipser-2000, albash-rmp-2018} to the implementation of effective imaginary time evolution \cite{motta-nature-2019, jouzdani-pra-2022}. All these methods present substantial drawbacks or limitations, including: dependency of the performance on the quality of the variational ansatz \cite{fedorov-materials-2022, lee-2022} and classical optimization to minimize the energy, in the case of variational methods; limitation of the convergence time that grows without bound for system size when phase transitions are encountered in the trajectories of the adiabatic processes; necessity for intermediate steps of tomography and costly classical post-processing for the imaginary time evolution technique. On present and near-term, noisy quantum computers, all of these problems pale in comparison to the issue of noise \cite{shaib-2023, urbanek-prl-2021, resch-karpuzcu-2021}. As current and near-term quantum devices are irreducibly noisy, controlling the strength and nature of the coupling of a quantum system to the environment is of paramount importance in the development of quantum technology. Though at a first glance counter-intuitive, coupling to the environment can actually yield substantial advantages when the system-bath couplings and bath structure are tuned appropriately \cite{kapit-review-2017, harrington-nature-phys-2022}. 

One might expect the coupling to the environment to increase the system's entropy, and thus to reduce its ``quantum efficiency." And indeed, the reduction of coupling to \textit{uncontrolled} degrees of freedom of the environment has enabled much of the progress in quantum information processing, particularly in solid state systems \cite{koch-pra-2007, manucharyan-science-2009, gyenis-prx-2021}. However, carefully engineering the bath coupled to a system in specific ways can effectively reduce the system's entropy. To this end, there are different approaches, including cases in which the bath is effectively very cold and thus acts as an entropy and/or energy ``dump" \cite{kapit-prx-2014, zaletel-prl-2021}, or the Maxwell's demon that is digital error correction codes \cite{maxwell-demon-vedral-2000, maxwell-demon-rex-2017}. Indeed, dissipatively stabilized states, and thermal states in particular, are a novel alternative to fault tolerant error correction in near term devices, with important theoretical and experimental developments in the last few years \cite{boykin-vrijen-2002, verstraete-wolf-cirac-2008, huelga-plenio-2013, geerlings-prl-2013, kimchi-schwartz-prl-2016, wang-scien-rep-2017, kaplan-roggero-2017, magnard-prl-2018, metcalf-prr-2020, polla-pra-2020, weimer-2020, feng-pra-2022, harrington-murch-2022, matthies-berg-2022, kishony-berg-2023, mi-rosenberg-2023, lambert-nori-2023, sannia-zambrini-2023, GoogleAI_april-2023, li-schuster_feb-2023, shtanko-movassagh-2021}.
Further, there is evidence that these states can be very difficult to reproduce classically \cite{troyer-prl-2005, shtanko-movassagh-2021, kapit-prr-2024}. 

Significant questions remain about the detailed structure of these states, and to find practical uses for them it is important to better understand them. An effective way to study them is within the frame of \textit{quantum thermodynamics}, an emerging field of physics that unites concepts of quantum information with concepts from thermodynamics such as entropy, work, heat, and temperature \cite{anders-2016, shabani-neven-google-2016, deffner-2019}. Some practical applications include, for instance, temperature estimation and improved metrology methods within the context of quantum thermometry using ``impurity probes" in many-body systems \cite{correa-prb-2018, stace-2018, rubio-prl-2021, alhambra-prx-2021, planella-prl-2022, beckey-prr-2022, brenes-pra-2023, mihailescu-pra-2023, mihailescu-2024}.

In this work, using extensive numerical simulations and theory, we explore dissipatively stabilized states within the framework of quantum thermodynamics, by applying our algorithm to two 1D many-body Hamiltonians, namely: the ferromagnetic Ising model with transverse field; and the Heisenberg model on a ring. The algorithm presented here is called the Relaxational Quantum Eigensolver (RQE) to be consistent with its first proposal in \cite{perez-phd-2021, kapit2021systems2}, and is formulated with near-term devices in mind. One of our goals here is resource efficiency, obtained by keeping the number of auxiliary ``shadow" qubits of the same order as the number of primary qubits. We first show that the steady state reached by our algorithm returns an average energy equal to a constant fraction of the ground state energy (assuming random states have energy zero), which steadily improves as the error rate decreases.
Then, we find that in many cases, dissipatively stabilized states obtained from our algorithm are well approximated by thermal states, with a temperature set by balancing error rates. In general, we do not know a priori that the dissipatively-stabilized states are thermal; indeed, undoubtedly there are cases in which they would not be. However, for the two 1D many-body Hamiltonians considered in  our work, it is a reasonable assumption, which we confirm through detailed simulations in this work. But even when the steady state of the system is well-approximated as thermal \cite{lipka-prl-2023, dutta-prb-2018, lewenstein-2023, shabani-neven-google-2016, foldager-2022}, the question of how to measure that temperature is nontrivial. This work presents three methods to characterize the temperature of the dissipatively-stabilized state obtained from our algorithm, one of which efficiently infers it from the populations of the qubits used as the cold bath. Moreover, this new thermometry technique is particularly resource efficient, as it requires only a small bath while also providing a method to estimate a system's temperature without the need of computationally-expensive methods based on exact diagonalization, or experimentally expensive processes such as full state tomography. In this work, we only look at condensed matter problems but the algorithm can be applied to disordered optimization problems as well.

In the remainder of this paper, we first describe the algorithm used to study the steady state of two many-body Hamiltonians of interest. Then, we present the results obtained to characterize the dissipatively-stabilized states, first showing the approximation of the ground-state energy and then presenting our thermometry studies. In doing so we benchmark our new method, quantify its accuracy and identify potential mechanisms that can cause it to report overly high temperatures (relative to the real effective temperature of the many-body state). Finally, we draw conclusions and offer an outlook for future work.

\section{RQE Algorithm Description}

To set up our RQE algorithm, we first choose a quantum many-body Hamiltonian $H_P$ that governs the primary qubits and that we wish to minimize. We then couple the primary qubits weakly to a second register of $N_S = N_P / 2$ ancillary, or ``shadow", qubits \cite{kapit-prx-2014, kapit-prl-2016, matthies-berg-2022, GoogleAI_april-2023}; the shadow qubits act as an engineered bath that removes excitations from the ``primary" system, acting as an approximate error correction mechanism. The choice of the number of shadow qubits $N_S$ was guided by the goal of resource efficiency; initial simulations yielded very similar results for $N_S = N_P$ and $N_S = N_P / 2$; therefore, we kept the latter, more resource-efficient value for $N_S$ throughout our work. The shadow qubit Hamiltonian $H_S$ is given by
\begin{equation}
\label{H_S}
H_S = \sum_{j = 1}^{N_S} \frac{\omega_{S_j}}{2} \sigma^z_{S_j} 
\end{equation}
with $\omega_{S_j}$ being the energy of the $j$-th shadow qubit; the coupling between primary and shadow qubits is described by the Hamiltonian 
\begin{equation}
\label{H_PS}
H_{PS}(t) = \sum_{j k} \Omega_{j k}(t) O_{P_j} O_{S_k}
\end{equation}
with $\Omega_{j k}(t)$ being the primary-shadow interaction energy coupling primary qubit $j$ to shadow qubit $k$, and $O_{P_j}$, $O_{S_k}$ Pauli operators acting on the $j$-th primary qubit and $k$-th shadow qubit, respectively. Unless otherwise specified, the Pauli operators for the primary-shadow interactions in this work are taken as $\sigma^y_{P_j} \sigma^y_{S_k}$. The total system Hamiltonian is $H_T (t) = H_P + H_S + H_{PS} (t)$.

In order to implement time evolution under the (total) Hamiltonian on a gate-model quantum computer, we must Trotterize it with a small timestep $dt$. If $dt$ is sufficiently small, though, we can approximate evolution by the so-called Trotter decomposition
\begin{equation}
\label{trotter-decompos}
e^{iH(t)dt} \ket{\psi} \approx \big( e^{iH_P dt} e^{iH_S dt} e^{iH_{PS}dt} + O(dt^2) \big) \ket{\psi}
\end{equation}
This Trotter decomposition allows us to implement a continuous time evolution Hamiltonian with a digitized, gate-based model. As we are focused on near term NISQ implementations, higher order operation approaches \cite{trotter-prx-2021} that can be implemented to mitigate Trotter error are counterproductive here due to gate error, and instead we use this simpler layering structure.

\subsection{RQE algorithm details} 

\begin{figure*}
\includegraphics[width=0.65\textwidth]{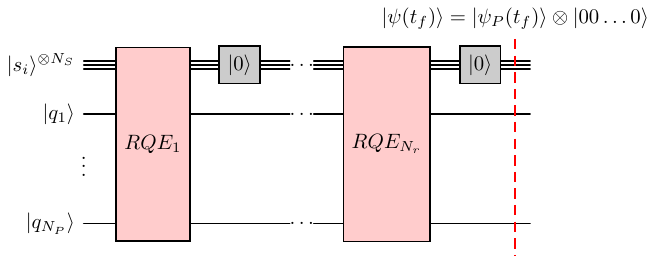}\\
\includegraphics[width=0.95\textwidth]{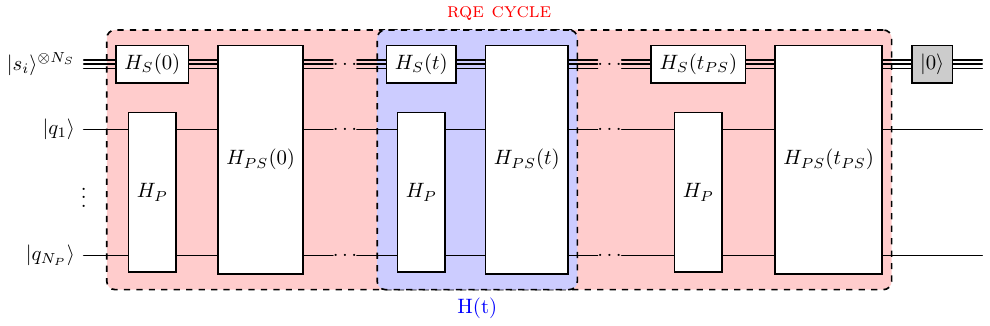}
\caption{\label{fig:algorithm-scheme}Schematic of the RQE algorithm (top) with high-level details of the algorithm implementation through Trotterized evolution (bottom): the entire bottom figure is inside each $RQE_n$ cycle. $H_P$ is the problem Hamiltonian acting on the primary qubit register, $H_S(t)$ acts on the shadow qubit register, and $H_{PS}(t)$ is the Hamiltonian governing the interaction between our primary and shadow qubit registers. We trotterize the total Hamiltonaian $H(t) = H_P + H_S(t) + H_{PS}(t)$ and evolve for some period of time before resetting the shadow qubits to the $|0\rangle$ state. We can tune terms in $H_S(t)$ and $H_{PS}(t)$ to optimize the transfer of energy and entropy from our primary qubits to our shadow qubits, which are subsequently reset to reduce the energy of the combined system.
}
\end{figure*}

A schematic of the RQE algorithm is shown in Figure \ref{fig:algorithm-scheme}: we take the initial state of the composite system as $\ket{\psi_{in}} \equiv \ket{\psi_{0, P}} \otimes \ket{0_S \cdots 0_S}$ with the shadow qubits prepared in the totally polarized state $\ket{0 \cdots 0}$. Then, at time $t=0$, the following sequence of gates is applied (we refer to this sequence as a \textit{single algorithmic layer})\cite{perez-phd-2021}:
\begin{enumerate}
\item Evolve the system under $\ket{\psi} \rightarrow e^{-2i\pi dt H_P} \ket{\psi}$, appropriately Trotterized; 
\item Apply the $Z$ rotations which define the shadow qubit Hamiltonian, $\ket{\psi} \rightarrow e^{-2i\pi dt H_S} \ket{\psi}$; this can be done in parallel with step (1); 
\item Apply the primary-shadow qubit interaction term, $\ket{\psi} \rightarrow e^{-2i\pi dt H_{PS}(t)} \ket{\psi}$; 
\item If $t$ is a multiple of $t_{PS}$, then reset all the shadow qubits to their ground state $\ket{0 \cdots 0}$; by periodically resetting the auxiliary qubits during the algorithm’s runtime we can effectively engineer an approximate zero-temperature bath which balances the effects of the infinite temperature bath from random gate error; 
\item If $t = t_f$, then halt and enact the appropriate gate sequence to measure $H_P$; otherwise, update $t \rightarrow t +dt$ and return to step (1).
\end{enumerate} 

\begin{figure*}
\includegraphics[width=0.85
\textwidth]{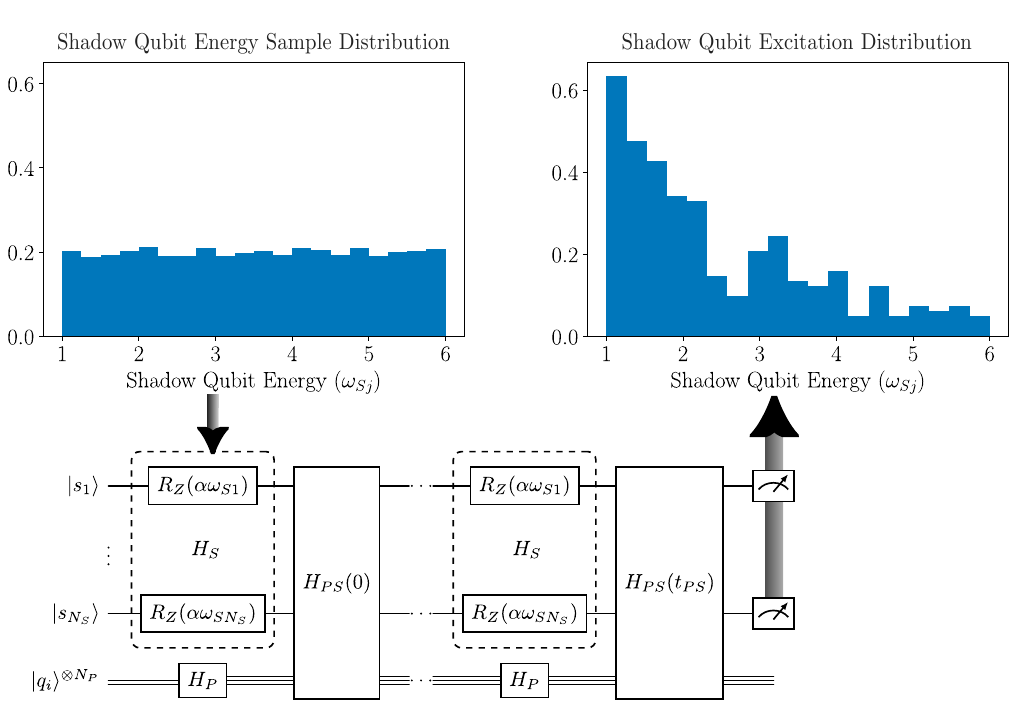}
\caption{\label{fig:thermometry-scheme}Schematic of the thermometry step where we sample shadow-qubit energies $\omega_{S j}$ from a uniform distribution (top left), run RQE, and measure the shadow qubits rather than resetting them (bottom). By repeating this we can extract a thermal distribution relating the probability of measuring an excited shadow qubit to its corresponding energy (top right).}
\end{figure*}

To study the thermal properties of the system, a thermometry step (shown schematically in Figure~\ref{fig:thermometry-scheme}) is implemented by performing a measurement at the end of the last reset cycle and getting the shadow qubit population for state $\ket{1}$ instead of resetting the shadow qubits to state $\ket{0 \cdots 0}$. This thermometry step is performed only at the end, not during the algorithm, because the shadow qubit energy is being varied in sweeps, which is more efficient for cooling, but makes it impossible to determine how much energy was extracted from the primary system when a $\ket{1}$ is measured. This step is used in the third method to evaluate the primary system's temperature by fitting the average temperature $\langle T \rangle$ from the Fermi-Dirac distribution, as explained below, and requires multiple time evolutions.
This thermometry step is key to the analysis of the dissipatively stabilized states obtained from our algorithm and we compare the thermometry results with two other methods used as benchmarks to characterize the temperature of the steady state of our RQE algorithm. The details of the three methods to estimate the temperature are explained in the next section.

\begin{figure}[h!]
\centering
\includegraphics[width=0.49\textwidth]{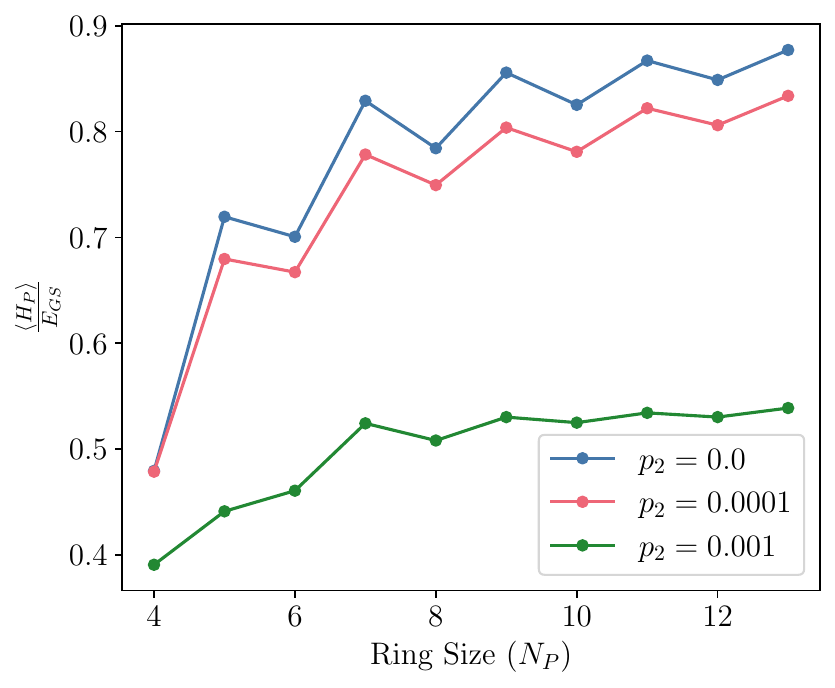}
\includegraphics[width=0.49\textwidth]{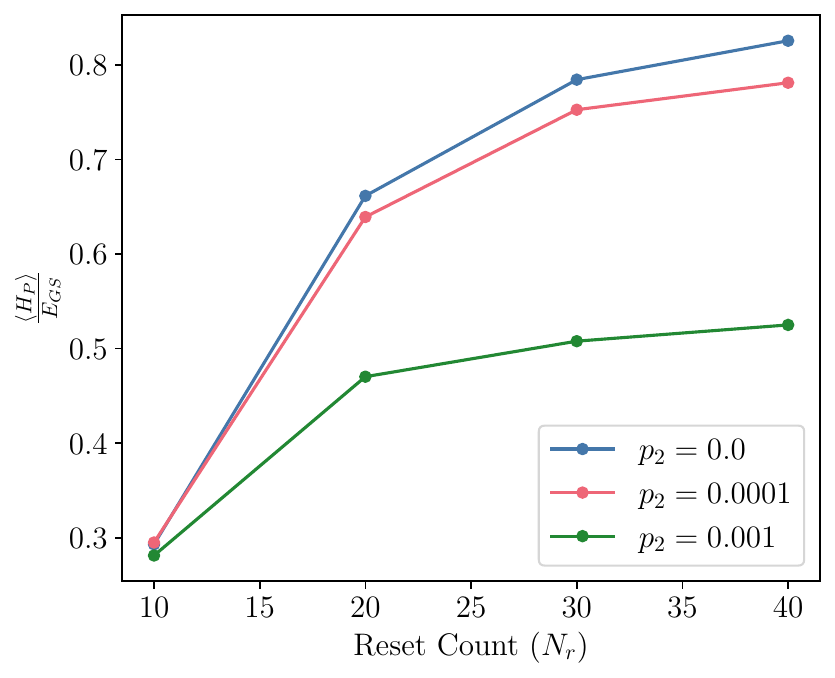}
\caption{\label{fig:approx_ratio} Ratio between the energy expectation value $\langle H_P \rangle$ for the steady state of the RQE algorithm and the energy of the ground state $E_{GS}$ of the antiferromagnetic Heisenberg chain, as a function of ring size $N_P$ for $N_r=40$ (left) and number of resets $N_r$ for $N_P=10$ (right), for different values of the two-qubit gate error rate $p_2$. For the problems studied in this work, random states have energy zero. The fact that the change in temperature is apparently small compared to the change in energy is to be expected because we are very close to the ground-state so the energy is exponentially sensitive to $1/T$; note that in FIG. 6 there is a slowly increasing temperature as well.}
\end{figure}

\begin{figure}[h!]
\centering
\includegraphics[width=0.45\textwidth]
{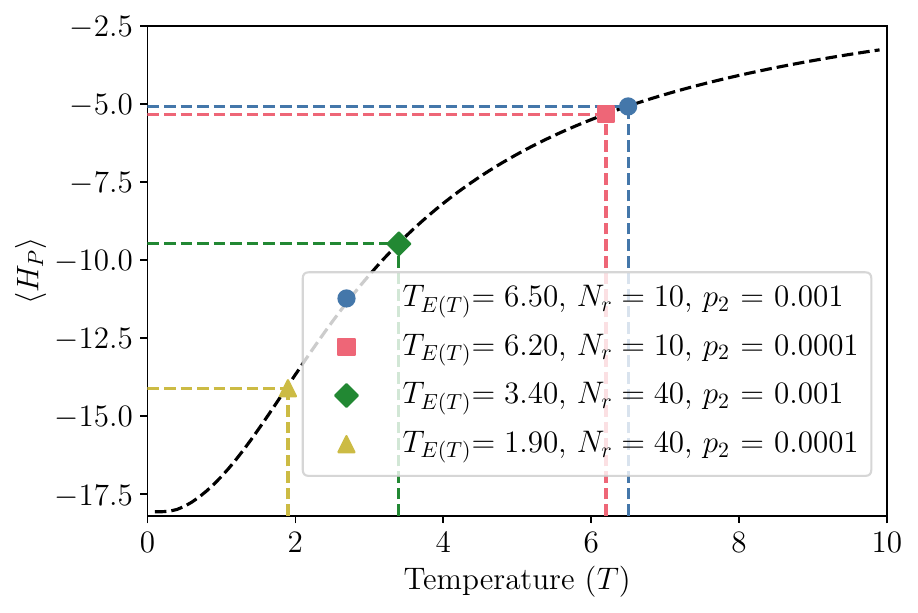}
\includegraphics[width=0.45\textwidth]
{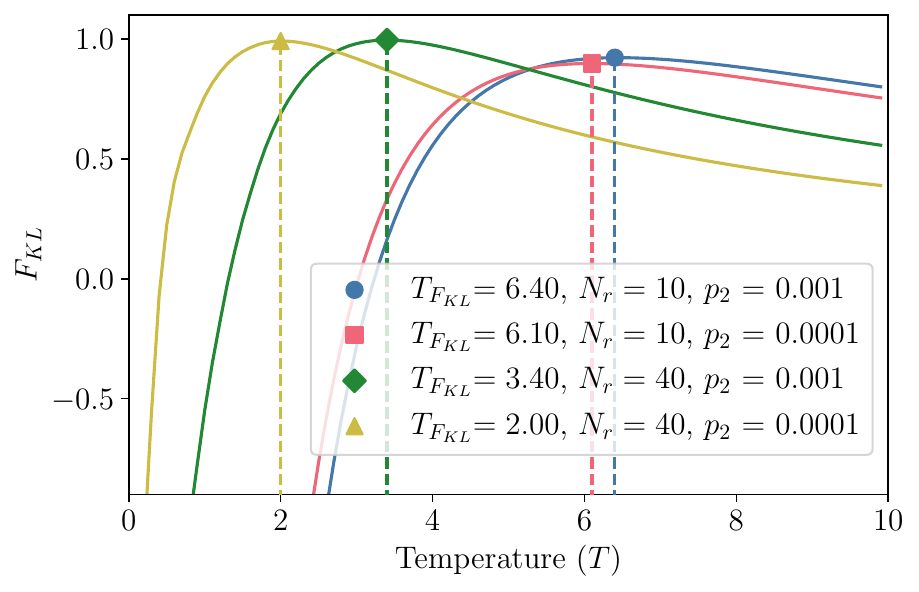}
\caption{\label{fig:Heisenberg-E(T)-nKLD} Thermometry of the dissipatively stabilized Heisenberg chain. Values of energy expectation values $\langle H_P \rangle$ as a function of temperature $T$ (left plot); fidelity measure $F_{KL}$ as a function of temperature $T$ (right plot); for both methods, the number of primary qubits is $N_P=10$ and we show results for two values of error rate $p_2 = 0.0001, \, 0.001$ and two circuit depths ($N_r = 10, \, 40$).}
\end{figure}

\vspace{1cm}

\begin{figure}[h!]
\centering
\includegraphics[width=0.32\textwidth]{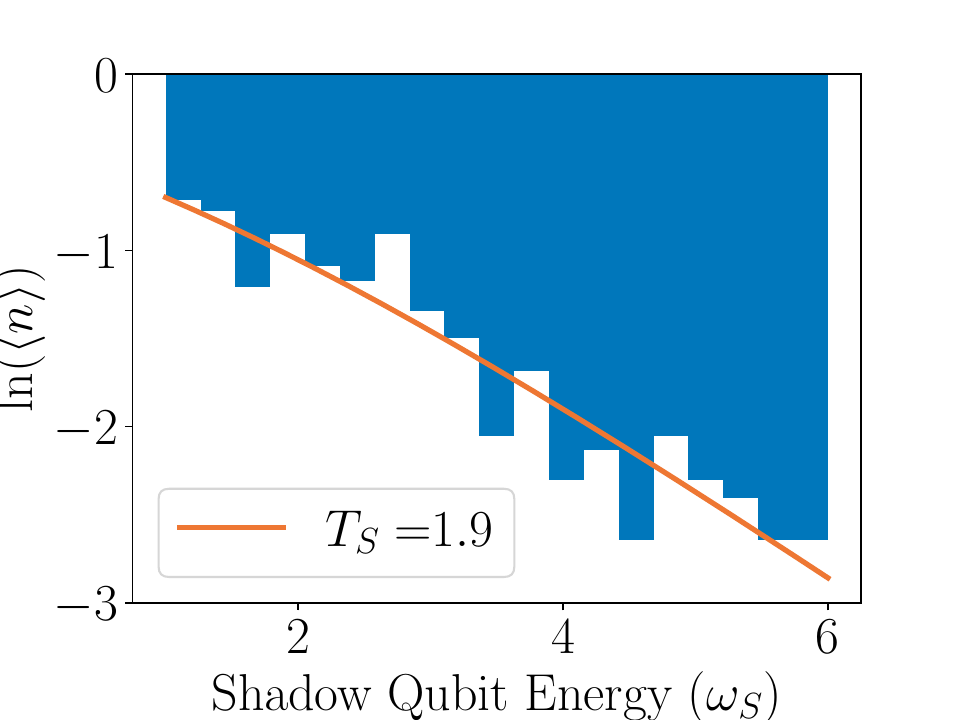}
\includegraphics[width=0.32\textwidth]{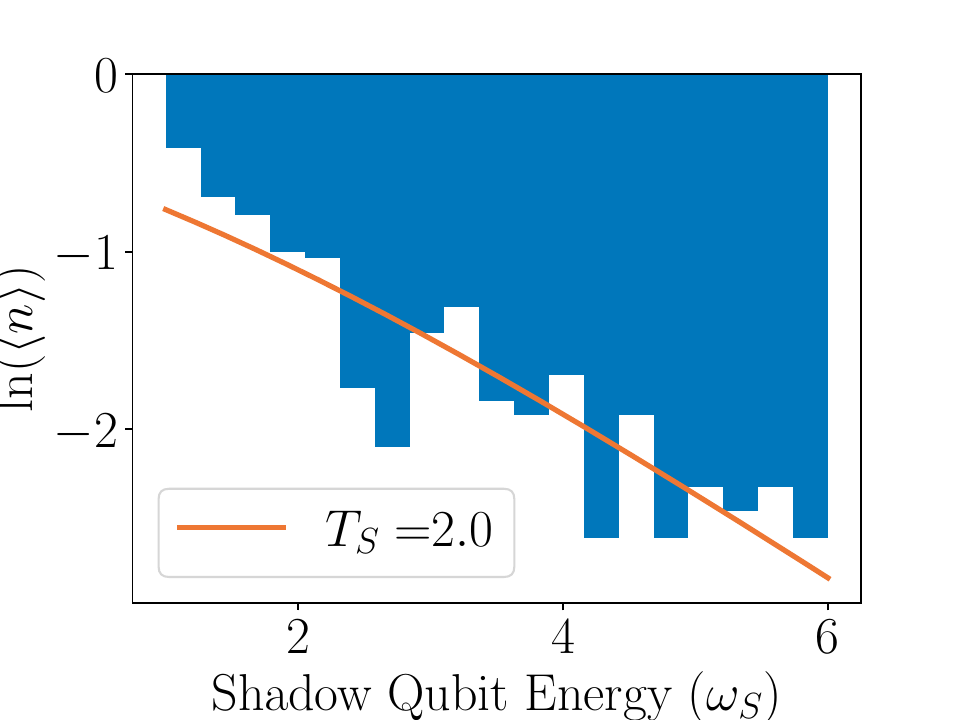}
\includegraphics[width=0.32\textwidth]{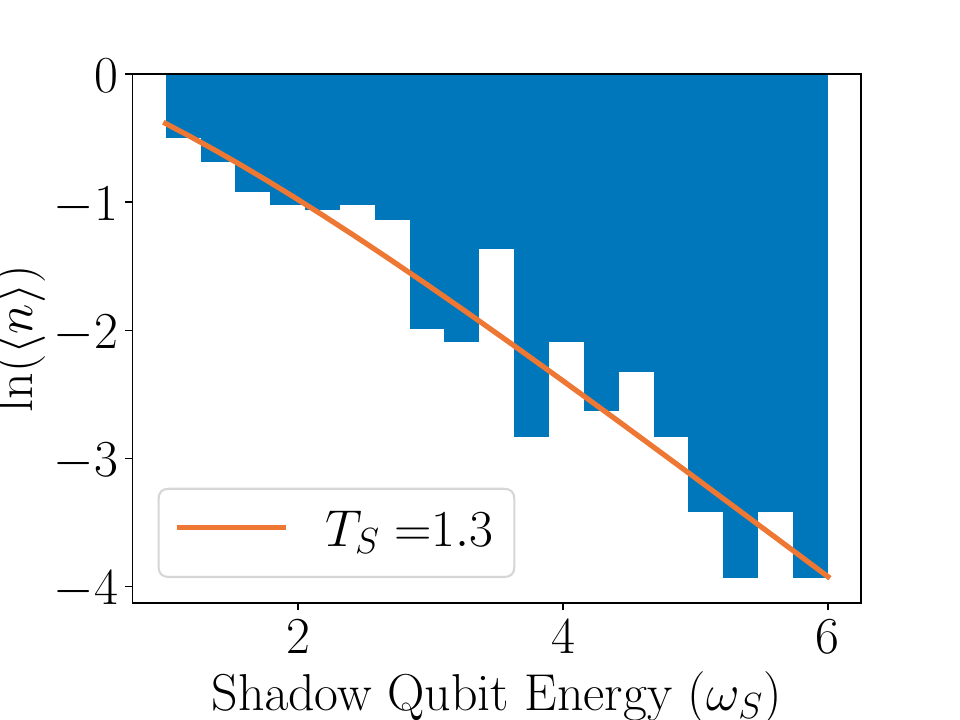}
\caption{\label{fig:thermal_dist} Logarithmically scaled shadow-qubit energy distributions, used to infer the shadow-qubit temperature $T_S$. Antiferromagnetic Heisenberg Chain (left); Ferromagnetic transverse field Ising model (TFIM) with $\kappa=0.75$ (center) and $\kappa=0.90$ (right); all with $N_P=10$, $N_r=40$, and $p_2 = 0.0001$.}
\end{figure}

The variational parameters of the RQE algorithm are:
\begin{itemize}
\item Algorithm time variable $t$ and global timestep $dt$; the latter is defined relative to the problem Hamiltonian: if in each timestep the system evolves via an appropriately Trotterized $\ket{\psi} \rightarrow e^{-i \alpha H_P}$, then $\alpha \equiv 2 \pi dt$. Note that $dt$ must be chosen carefully to balance Trotter error and gate error, so in this work, we chose $dt$ to be small enough to eliminate Trotter error while large enough to be able to meaningfully evolve the system \cite{schuster-2023}; 
\item Runtime of the algorithm $t_f$ which defines the number of algorithmic layers of gates $N_L \equiv t_f / dt$;
\item Time duration of a primary-shadow interaction pulse, $t_{PS}$, in which the primary-shadow interaction is ramped up and down before the shadow qubits are measured or reset; this defines the number of algorithm layers per pulse $N_{L,PS} \equiv t_{PS} / dt$. For the cycles per reset in this work, after trying a few different options, we found that for our given $dt$, 20 was a reasonable value yielding a sufficiently long evolution allowing for the system to thermalize without proliferating error dominating the system. In comparison to other work \cite{GoogleAI_april-2023}, we use longer cycles, the coupling in our algorithm is ramped up and down (instead of kept constant), and we focus on slightly lower levels of simulated noise;
\item Number of times the shadow qubits are reset to their ground state, $N_r$;
\item Shadow-qubits energies $\omega_{S_j}$: in our normal RQE algorithm, they are chosen similarly to \cite{matthies-berg-2022}, performing a downhill sweep for these energy values from 6 to 0.75 for $3/4$ of the evolution time and then holding the values at 0.75 for the remainder of the time. However, in a Thermometry step as seen in Figure \ref{fig:thermometry-scheme}, we randomly sample shadow-qubit energies from a uniform distribution, $\omega_{S j} \sim U(1,6)$, enabling us to extract the distributions shown in Figures \ref{fig:thermometry-scheme} \& \ref{fig:thermal_dist};
\item Primary-shadow interaction energies $\Omega_{j k}(t)$: for this interaction energy coupling primary-qubit $j$ to shadow-qubit $k$ we use a function that smoothly ramps up and down, specifically implementing a dome function $\Omega_{PS}(t) \equiv 4t(1-t)$, where 
$t\equiv (k + 1) / (N_{LPS} +1)$
with $k$ being the Trotter step index before reset; in our simulations we take this dome function to have a phase sum (i.e. area under the curve) of 
$\sum_{t=0}^{t_{PS}} \Omega_{PS}(t)dt = \pi / 4$; 
\item Error rate per operation $p_j$: generally, this includes single-qubit error rate $p_1$, two-qubit gate error rate $p_2$, measurement error rate $p_M$, and reset error rate $p_R$. Typically $p_1 \approx p_2 / 10$ as a phenomenological model \footnote{More structured error models, such as the loss/dephasing model in transmon qubits, can potentially be corrected more efficiently.}; $p_2$ represents the error per composite two-qubit unitary, e.g. $e^{i \gamma \sigma^z_j \sigma^z_k}$. In particular, in our algorithm we simulate depolarizing noise by randomly choosing one qubit for each 2-qubit gate appended and applying one of the Pauli matrices with probability $p_2 / 3$.
\end{itemize}

The parameter choices of our algorithm balance algorithmic efficiency and efficacy. The choices of $dt$, $N_{L,PS}$, and $N_r$ are directly related to circuit depth, the simulation runtime, and Trotter error. The initial values of many parameters were intuitively guided and later refined through variational methods to achieve the results presented in this work. One of the main key factors was finding the balance between the Trotter timestep $dt$ and the number of layers in an RQE cycle $N_{L,PS}$ to minimize Trotter error while minimizing the expectation value of the prepared state. In our work we found that $dt=0.0667$ and $N_{L,PS}=20$ balanced these criteria fairly well, though obviously the best choice depends on the problem Hamiltonian $H_P$, base error rate and other details. Additionally, our parameter choices for the shadow qubits and reset rates dictate how well our engineered bath can cool the system. Similarly, the initial choices of the shadow qubit energies $\omega_{S_j}$, reset count $N_r$, and coupling strengths $\Omega_{jk}(t)$ were theoretically guided and later varied during experimentation to optimize algorithmic efficacy.

\section{Ground state approximation}

We first consider the degree to which RQE is able to approximate the ground states of these systems. The first system we consider is a ring of $N_P$ primary qubits in a 1D ferromagnetic Ising model with transverse field, described by the primary Hamiltonian $H_P$
\begin{equation}
\label{H_TFIM}
H_P = -J \sum_{j = 1}^{N_P} \sigma^z_j \sigma^z_{j+1} -J \sigma^z_1 \sigma^z_{N_P} - \kappa \sum_{j = 1}^{N_P} \sigma^x_j 
\end{equation}
with ferromagnetic energy scale $J$ (set to 1 for simplicity) and transverse field strength $\kappa < 1$. \\ 
The second system we consider is a ring of $N_P$ primary qubits in an antiferromagnetic Heisenberg model, described by the primary Hamiltonian
\begin{eqnarray}
 \label{H_Heis}
 H_P = J \sum_{j = 1}^{N_P} \big( \sigma^x_j \sigma^x_{j+1}
 + \sigma^y_j \sigma^y_{j+1} + \sigma^z_j \sigma^z_{j+1} \big) + 
 J \big( \sigma^x_1 \sigma^x_{N_P} + \sigma^y_1 \sigma^y_{N_P} + \sigma^z_1 \sigma^z_{N_P} 
 \big).
 \end{eqnarray}
`'
In this work, we set the initial state of the primary system $\ket{\psi_{0, P}}$ as an antiferromagnet for the ferromagnetic systems (TFIM) and for the antiferromagnetic systems (Heisenberg) we had the initial state as a ferromagnet. These choices were made because they allowed us to watch RQE actually work rather than starting in an initial state that was already close to the ground state.
 
In Figure~\ref{fig:approx_ratio} the ratio between the energy expectation value $\langle H_P \rangle$ for the steady state of the RQE algorithm and the energy of the ground state $E_{GS}$ of the 1D antiferromagnetic Heisenberg ring is plotted as a function of the number of primary qubits $N_P$ (left) and as a function of the number of resets $N_r$ (right). In general, for a random state the energy is zero. The energy expectation value of the steady state of the RQE algorithm approximates the ground-state energy with results improving for decreasing error rate $p_2$, as expected. These results align well with other recent works \cite{matthies-berg-2022, GoogleAI_april-2023}, and improve both with increasing number of primary qubits $N_P$ and with higher-depth circuits (i.e. increasing number of resets $N_r$). 

To understand why RQE both returns a constant fraction approximation of the ground state energy, and does not reach the exact ground state for zero error (for the parameters chosen), let us consider the algorithm as an approximation to continuous time evolution. Random gate error introduces an error rate $\Gamma_P$ that heats the system toward infinite temperature; the primary-shadow interaction increases this to $\Gamma_P \left ( 1 + \epsilon \right ) + \Gamma_R'$, where the factor of $\epsilon$ comes from increasing circuit complexity per timestep and $\Gamma_R'$ is a small off-resonant heating rate from the shadow qubits themselves, which vanishes as $\Omega_{PS} \to 0$ but is nonzero for the parameters chosen here (more on this below). Balancing this rate, the shadow qubits induce an average excitation removal or cooling rate $\Gamma_R$ which depends on the algorithm parameters and not gate error. Hence, with appropriate tuning, we can obtain a substantial cooling rate $\Gamma_R \gg \Gamma_P$ which should allow our primary system to eventually thermalize and approximate the ground state of the corresponding many-body Hamiltonian $H_P$. Maximizing $\Gamma_R$ is thus key and can be achieved by leveraging the resonant transitions by varying the shadow-qubit energies $\omega_{S j}$ during the algorithm's runtime. We can thus expect the long-time residual energy to scale roughly as $\Delta E \approx N_P \frac{\Gamma_P(1 + \epsilon) + \Gamma_R'}{\Gamma_R}$ and therefore the energy of the steady state of the RQE algorithm to approximate the ground-state energy to a constant fraction. In general, for some systems such as quantum spin glasses, the time to equilibrate may be prohibitively long \cite{garrahan-prb-2015, king-nature-2023}; it is empirically fairly short here but this is not a universal property of the algorithm.

\begin{figure}[h!]
\centering
\includegraphics[width=0.49\textwidth]{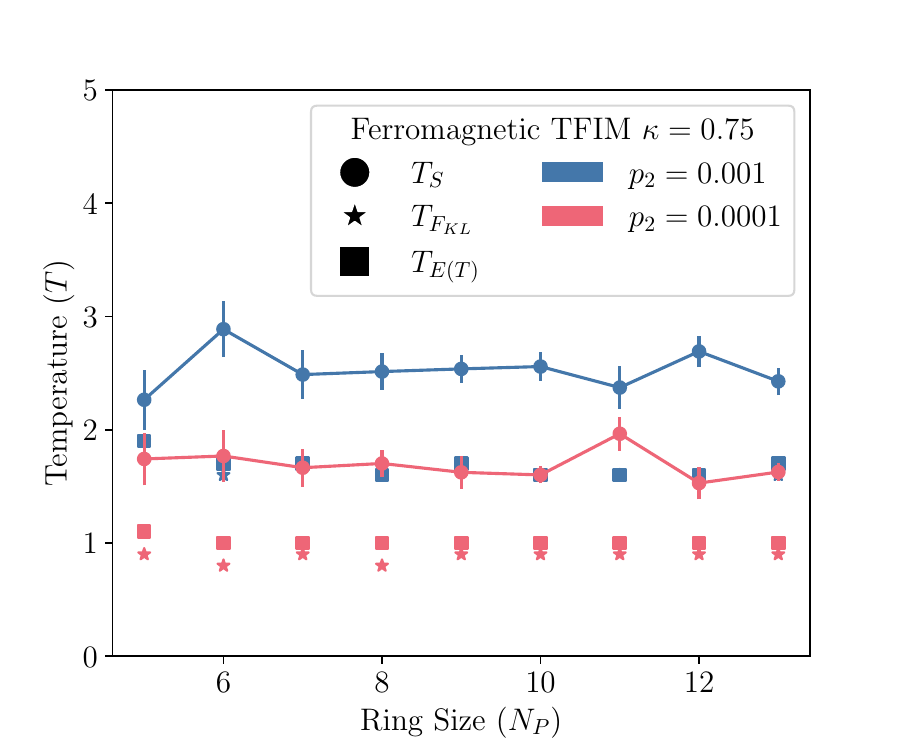}
\includegraphics[width=0.49\textwidth]{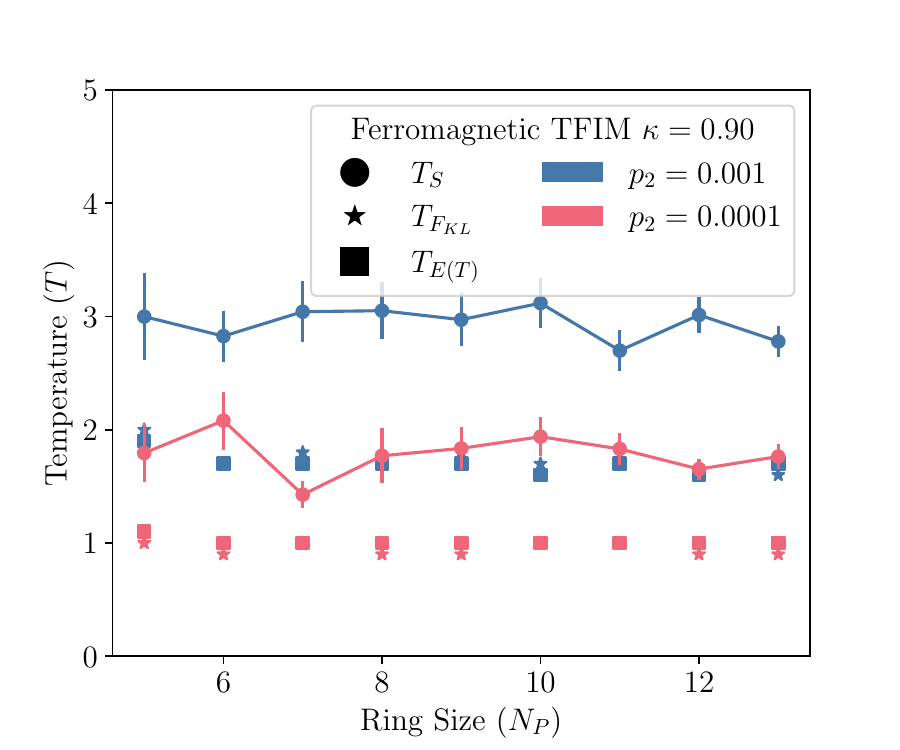}\\
\includegraphics[width=0.49\textwidth]{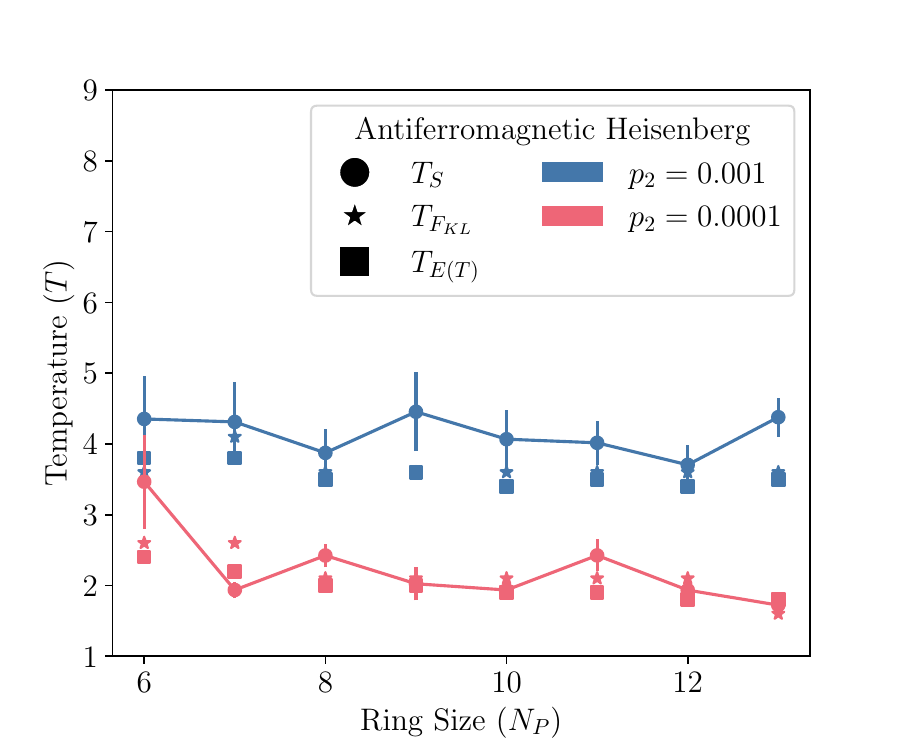}
\includegraphics[width=0.49\textwidth]{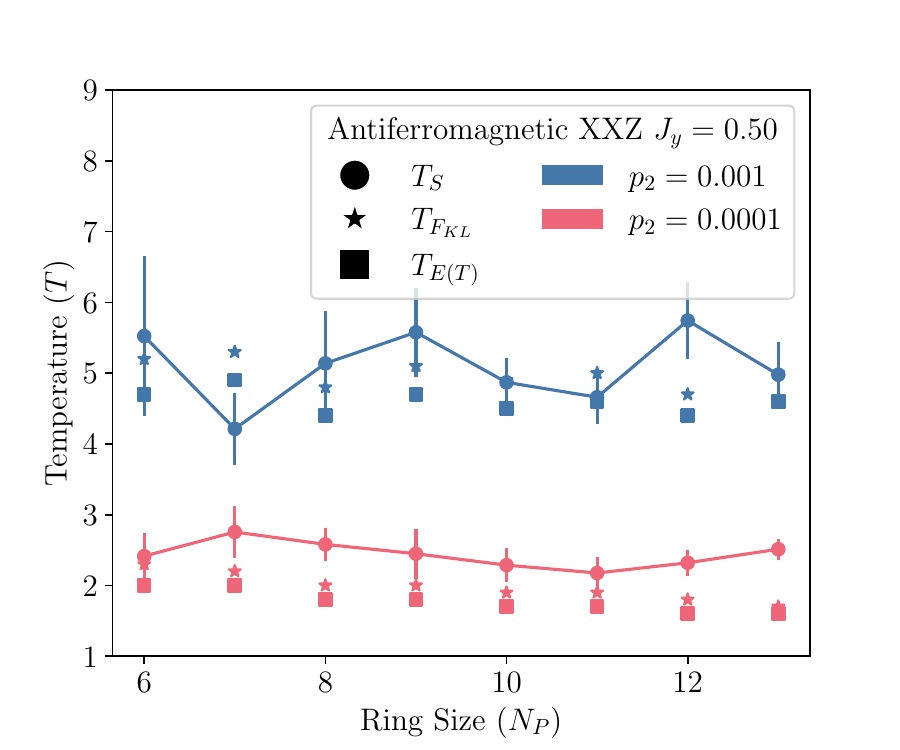}
\caption{\label{fig:temp_comp} 
 Temperatures extracted from the three different methods (explained in Section IV) plotted versus system size: $T_{F_{KL}}$ from the fidelity measure $F_{KL}$; $T(E)$ from the energy expectation values; $T_S$ from the shadow qubit populations. Ferromagnetic Transverse Field Ising Model with $\kappa = 0.75$ (top left) and $\kappa = 0.90$ (top right); Antiferromagnetic Heisenberg ring (bottom left); Antiferromagnetic XXZ with primary-bath operator $O_{P_j} O_{S_k} \equiv \sigma^y_{P_j}\sigma^y_{S_k}$ and $J_y=0.5$ (bottom right).  All four systems are plotted for high depth circuits, i.e. $N_r$ = 40, with time step $dt=0.0667$ and number of algorithm layers per pulse $N_{L,PS}=20$; the primary-shadow qubit coupling is $2:1$ and the shadow-qubit energies are randomly sampled from a uniform distribution, $\omega_{S_j} \sim U(1,6)$. The data is from 1000 random trajectories per point; the error bars are one standard deviation of the temperature values from sampling and fitting the shadow qubit distributions.}\label{mainTfig}
\end{figure}

\begin{figure}[h!]
    \centering
    \includegraphics[width = 0.49\textwidth]{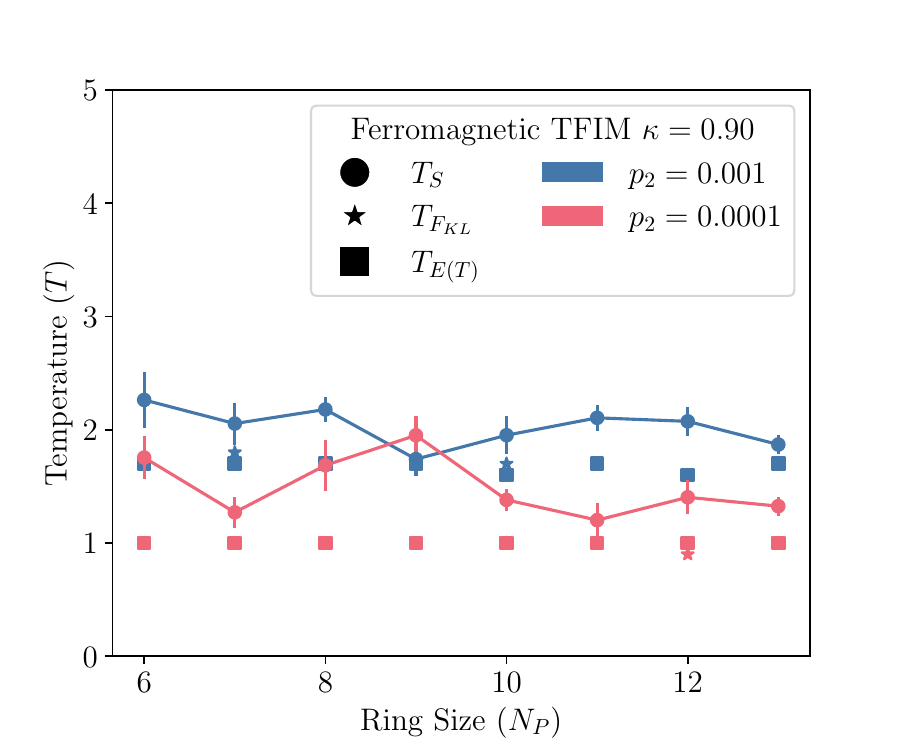}
    \includegraphics[width=0.49\textwidth]{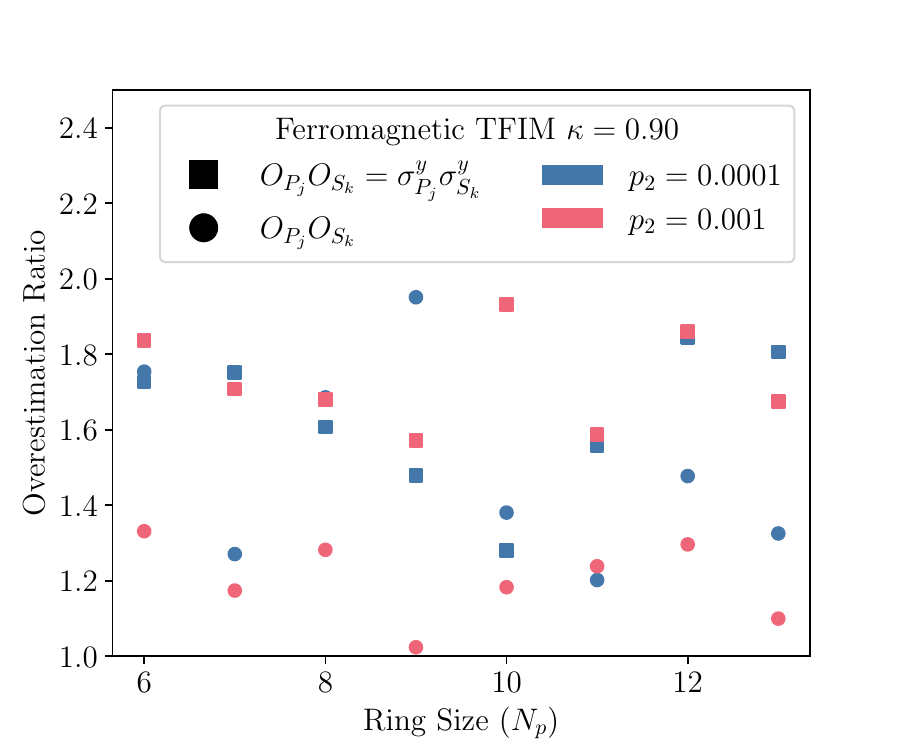}
    \caption{Temperature extracted using the three different methods (i.e., $T$ from the energy expectation value $\langle H_P \rangle$; $T$ from the distribution that minimizes the fidelity measure $F_{KL}$ in the $Z$-basis; $T$ from the shadow qubit populations) plotted versus system size: Here we slightly modify the thermometry method by randomly selecting the primary-bath operator $O_{P_j} O_{S_k}$ from $\{\sigma^x_{P_j}\sigma^x_{S_k},\sigma^y_{P_j}\sigma^y_{S_k},\sigma^z_{P_j}\sigma^z_{S_k}\}$ and averaging over multiple samples (left); Temperature overestimation ratio, $2T_S / (T_{F_{KL}} + T_{E(T)})$, plotted versus system size comparing our traditional primary-bath operator $O_{P_j} O_{S_k} \equiv \sigma^y_{P_j}\sigma^y_{S_k}$ to the randomly sampled operator (right). Both plots are for TFIM with $\kappa=0.90$ and $N_r = 40$. The error bars are one standard deviation of the temperature values from sampling and fitting the shadow qubit distributions. The three thermometry methods as well as the overestimation are discussed in Section IV.}
    \label{fig:random_bath}
\end{figure}

\section{Thermometry}
 
We now turn to the thermometry of these dissipatively stabilized states. In order to better understand the thermal behavior of the steady state obtained from the RQE algorithm, we use three methods to characterize the temperature of the dissipatively stabilized state, namely:
\begin{enumerate}
\item Inferring $T$ from the energy expectation value $\langle H_P \rangle$;
\item Inferring $T$ from the distribution that minimizes \textit{K-L divergence} between the true thermal density matrix and measurements of bitstrings in the $Z$-basis;
\item Inferring $T$ from the shadow qubit populations.
\end{enumerate} 

While the first two methods are both based on exact diagonalization of the primary system Hamiltonian, the third method is applicable on real quantum hardware and to cases where exact diagonalization is not possible. The first two methods are used to calibrate our expectations and check the results obtained from our RQE algorithm with the third method: in fact, as shown in Figure~\ref{fig:temp_comp}, the temperatures extracted from the three different methods are broadly consistent with each other.

For the first method, in  particular, we are assuming that the steady state is thermal: while in general this is not always the case, for the two 1D many-body Hamiltonians considered in  our work, it is a reasonable assumption, which is also confirmed by the results obtained with the K-L divergence method presented in the next section. 
Thus, we can use an \textit{operational definition of temperature} (as in \cite{shabani-neven-google-2016}), consider the system consisting only of primary qubits, and fully diagonalize it. From the exact diagonalization, we obtain $2^{N_P}$ energy eigenvalues $\{ E_j \}_{j=1,\dots,2^{N_P}}$. Then, for each temperature \footnote{In order to simplify the notation and calculations, we have set $k_B=1$ (and $\hbar=1$) so our temperature here has the same units as energy.} value $T$ in a given range, we compute the partition function $Z = \sum_{j=1}^{2^{N_P}}e^{-E_j/T}$ and the thermal expectation value for the energy as a function of the temperature $T$: 
\begin{equation}
\label{E(T)}
E(T) = \frac{\sum_{j=1}^{2^{N_P}} E_j e^{-E_j/T}}{Z} 
\end{equation}
Hence we extrapolate at what specific temperature value $T^*$ the curve $E(T)$ crosses the energy expectation value $\langle H_P \rangle$ obtained from the RQE algorithm. Moreover, we can also use this result from the exact diagonalization to compare it with the temperature inferred from the shadow-qubit populations, as explained in the third method below.

For the second method, we start by measuring the distance between the final state obtained from the RQE algorithm and the thermal distribution obtained from the transverse-field Ising model with a fixed number of primary qubits on a ring. We measure this distance with the Kullback-Leibler divergence (a.k.a. \textit{K-L divergence} or \textit{relative entropy}), with which one can measure how close, or similar, one statistical distribution $P$ is to a given, or reference, statistical distribution $Q$ \cite{csiszar, kullback-leibler}. 
In our specific case, the distributions are discrete; the given, or reference, distribution $Q$ can be taken to be the thermal state obtained from exact diagonalization of the primary qubits Hamiltonian; thus, the $Z$-basis distribution obtained from the RQE algorithm is the distribution $P$ for which we want to calculate the distance from the reference distribution $Q$. \\
We use the standard definition of the K-L divergence: 
\begin{equation}
\label{K-L divergence-def}
D_{KL}(P \; || \; Q) \equiv \sum_{j=1, \dots, 2^{N_P}} = P(j) \log \bigg( \frac{P(j)}{Q(j)} \bigg) 
\end{equation}
From this definition, as in~\cite{kapit-prr-2024}, we want to find the temperature (Figure ~\ref{fig:Heisenberg-E(T)-nKLD}) which maximizes the normalized fidelity measure
\begin{equation}
\label{norm-kld}
F_{KL} \equiv 1 - \frac{D_{KL}(\rho_{therm} || \rho_{RQE})}
{D_{KL}(\rho_{IUR} || \rho_{RQE})}
\end{equation}
where $\rho_{therm} = e^{-H/T}/\text{Tr}(e^{-H/T})$ is the ideal thermal state, $\rho_{RQE}$ is the dissipatively stabilized state obtained from our algorithm, and $\rho_{IUR}$ is the random uniform distribution where all strings have probability $2^{-N_P}$. The normalization of the fidelity measure $F_{KL}$ is performed by the incoherent uniform random distribution since this is the distribution we would get in a high depth circuit with no error correction. We use this method as a second check of the system's temperature since it is closer to what can be done on hardware, though it is expensive as it requires many samples to converge (and requires us to calculate $\rho_{therm}$ for comparison). However, it is much more efficient than full state tomography once the system size is large. As shown in Figure~\ref{fig:Heisenberg-E(T)-nKLD}, the temperatures extracted from this method and from the average energy show excellent quantitative agreement, confirming the thermal character of these states.

Finally, with the third method we estimate the temperature of the steady state obtained from the RQE algorithm by fitting the average temperature $\langle T \rangle$ from the number of shadow qubits in the excited state $\ket{1}$ using the Fermi-Dirac distribution 
\begin{equation}
\label{Fermi-Dirac-distrib}
\langle n\rangle = \frac{e^{- \omega_{S_j}/T}}{1 + e^{- \omega_{S_j}/T}}
\end{equation}
Since the shadow qubits act as a bath draining energy from the system of primary qubits by extracting excitations, the number of shadow qubits found in the state $\ket{1}$ at the end of the algorithm can be used as a measure for the temperature of the primary system (given fixed shadow qubit $\omega_{S_j}$). 

We apply each of the three methods both to the 1D ferromagnetic Ising model with transverse field and to the antiferromagnetic Heisenberg ring. For both models, we performed simulations with different circuits depths (i.e. with the number of resets ranging from $N_r=10$ up to $N_r=40$), for error rates $p_2 = 0, \, 0.0001, \, 0.001$, and for rings with a varying number of primary qubits $N_P = [4, \dots, 13]$. For the Ising model, we considered two different values of transverse field strength $\kappa = 0.75, \, 0.9$.

 In Figure~\ref{fig:Heisenberg-E(T)-nKLD} we show results from the first two (i.e. the benchmark) methods for the Heisenberg model. As can be seen here, $F_{KL}$ as defined in (\ref{norm-kld}) reaches its maximum value close to unity proving that the state reached with the RQE algorithm in fact approaches the thermal distribution. Moreover, the temperature for which the normalized K-L divergence reaches its maximum value is consistent with the temperature $T(E)$ obtained from the energy expectation values. It is important to note that for the TFIM the temperature does not reach an exactly zero value. One of the physical causes of this is the nature of the excitations in the system. In fact, the elementary excitations in the ferromagnetic phase are non-local and topological, since they are domain walls, and as such they can only be created and destroyed in pairs, in the bulk \cite{matthies-berg-2022, kishony-berg-2023, lewenstein-2023}. Thus, topological excitations cannot be removed by local operations; therefore, two or more of such excitations need to coincide in order to be removed. This makes it effectively harder to cool quantum systems with non-local (topological) excitations and challenging to prepare topological ground states on quantum simulators.

For the third method, we have performed simulations for different ring sizes, with primary qubit range $N_P = [4, \dots, 13]$, varying the number of resets $N_r$ of the shadow qubits and error rates $p_2 = 0.0001, \, 0.001$. 
In Figure \ref{fig:thermal_dist} we show the logarithmically scaled shadow-qubit energy distributions, used to infer the shadow-qubit temperature $T_S$ for three cases, all with $N_P=10$, $N_r=40$, and $p_2 = 0.0001$: antiferromagnetic Heisenberg ring (left); ferromagnetic transverse field Ising model (TFIM) with $\kappa=0.75$ (center) and $\kappa=0.90$ (right). These extrapolated temperature values are fairly consistent with those obtained from the other two methods, as shown also in Figure~\ref{fig:temp_comp}. In particular, for the 1D ferromagnetic Ising model with transverse field we see a relatively constant overestimation of the thermometry temperature, $T_S$, which remains consistent with fixed error rates over varying circuit depths; whereas for the antiferromagnetic Heisenberg ring with $N_P > 6$ we get more accurate predictions of the thermometry temperature, $T_S$, improving as system size increases. 

An intuitive explanation for the temperature overestimation can be given based on concepts from \cite{kapit-review-2017}. As the interaction strength isn't strictly zero, the probability of the primary-shadow interaction inducing ``off resonant" transitions (e.g. interactions where $dE_P + dE_S \neq 0$) is also non-zero. The rate $\Gamma_{\pm}(\Delta)$ of those processes is suppressed only at low-order polynomially in the energy mismatch $\Delta$ and it is symmetric about $\Delta = 0$ \cite{kapit-review-2017}: 
\begin{equation}
\label{transit-rate}
\Gamma_{\pm}(\Delta) = \frac{ \Omega_{jk}^2 \omega_{S_j}/2} {(\Omega_{jk}^2 \pm \Delta)^2 + \omega_{S_j}^2 /16}
\end{equation}
In contrast, the expected shadow-qubit population $\langle n\rangle$ vs. $E$ (as a function of temperature) is an exponential distribution; therefore, if we add some ``uncertainty" from such resonance effects, then that blurring is going to symmetrically increase the expected population above a given energy just as much as it increases the population below it, which in turn is going to artificially increase $T$ in fitting. Furthermore, gate error which excites a shadow qubit will lead to a higher temperature measurement (though with $p_1 = p_2/10$ this will be rare). Thus, on balance we expect that this method will tend to overestimate the temperature.

This argument does not explain however why the shadow-qubit estimated temperatures display a much larger relative overestimation for the TFIM than in our Heisenberg simulations (see Figure~\ref{fig:temp_comp}). While we expect the topological character of elementary excitations plays a role here, it turns out the choice of operator ($Y_j$ by default) used for the system-bath coupling also contributes significantly to the overestimation. Specifically, the TFIM in our notation is composed of $ZZ$ interactions with a transverse field along $X$, whereas the Heisenberg chain is isotropic and its Hamiltonian does not have any ``preferred directions" in operator space. While given an infinite time the primary and shadow qubits should perfectly thermalize (assuming, of course, the primary system is thermal in the first place), our thermometry method only collects excitations for a single cycle. It stands to reason that if particular operator choices bias the transfer of excitations (as a function of energy or any other parameter), then this may contribute significantly to ``error" in thermometry from short time measures such as ours.

In order to address the difference in the accuracy of the temperature extrapolated from the thermometry method for the TFIM vs. Heisenberg model, we performed simulations for the XXZ model keeping all parameters the same as for the Heisenberg ring except for decreasing the exchange energy between nearest-neighbors in the $y$ direction, i.e. $J_x = J_z = 1$ but $J_y = 0.5$. In this case, as seen in the lower right panel of Figure~\ref{fig:temp_comp}, the accuracy of the temperature extrapolated from the thermometry method is not as close as the Heisenberg model but better than the TFIM. And in order to further confirm these expectations, we performed simulations for the TFIM with $\kappa=0.9$ randomly sampling the primary-shadow interaction operators $O_{P_j} O_{S_k}$ from $\{\sigma^x_{P_j}\sigma^x_{S_k},\sigma^y_{P_j}\sigma^y_{S_k},\sigma^z_{P_j}\sigma^z_{S_k}\}$, as shown in Figure~\ref{fig:random_bath}: as can be clearly seen here, the estimation of the temperature inferred from the RQE thermometry method improves when randomly sampling the primary-shadow interaction operators $O_{P_j} O_{S_k}$ from $\{\sigma^x_{P_j}\sigma^x_{S_k},\sigma^y_{P_j}\sigma^y_{S_k},\sigma^z_{P_j}\sigma^z_{S_k}\}$ vs. $O_{P_j} O_{S_k} \equiv \sigma^y_{P_j} \sigma^y_{S_k}$. 

We present all these simulations to note and carefully illustrate how operator choice is a potential source of error in thermometry through this method. We note, of course, that the choice of primary-shadow interaction operator is another variational parameter for our algorithm, and the choice which is best for cooling the primary qubits (we chose $Y_j$ because it led to the lowest energies in our early simulations used to set parameters) may not be the best for accurately measuring the temperature using the dissipative elements.

\section{Conclusions \& Outlook}

In this work, we characterized the dissipatively stabilized states of our relaxational quantum eigensolver (RQE) algorithm by estimating their temperature using three different methods which yield broadly consistent results both for the 1D ferromagnetic Ising model with transverse field and for the antiferromagnetic Heisenberg ring. The first two methods, minimizing the normalized K-L divergence between the distribution obtained by $Z$ basis measurements and a thermal average, and estimating temperature from the mean energy, require exact diagonalization of the system (or similar computationally expensive methods) to create the ideal distributions to which our dissipative algorithm results are compared. They are thus not suitable for acheiving advantage in quantum simulation, but are a useful tool to characterize these states at small scales and set expectations, and both measures returned temperatures that were very close to each other (and likely within sampling error) in all cases, showing that the steady states of these two problems are indeed thermal to good approximation.

To provide a more efficient measure of the temperature, we introduced a thermometry method based on the dissipative elements themselves (``shadow qubits" in the language of our algorithm). We showed that while this method tends to overestimate the temperature on general grounds, and care must be taken in operator choices to reduce this effect, it nonetheless can provide qualitatively and often quantitatively good thermometry results (using the first two methods as a benchmark). This thermometry method is simple and scalable, and could be implemented in real quantum hardware to measure the temperature of dissipatively stabilized steady states at beyond classical scales. For instance, it could be applied to non-stoquastic primary Hamiltonians $H_P$ (where Quantum Monte Carlo has a sign problem), where getting thermal states can be classically hard. Moreover, given that this new thermometry technique uses only $O(N_P)$ shadow qubits (i.e. a small external bath), one of its key values is resource efficiency; it represents a novel use of the quantum resources of the system by providing an approximate error correction mechanism that can be used for other purposes too. 

Moreover, we also calculated the ratio between the energy expectation value $\langle H_P \rangle$ for the steady state of the RQE algorithm and the energy of the ground state $E_{GS}$ for the models studied, showing that the energy of the steady state of the RQE algorithm reaches a constant fraction of the ground-state energy, with results improving for decreasing error rate $p_2$, as expected, and in good agreement previous works \cite{matthies-berg-2022, GoogleAI_april-2023}. Our results improve both with increasing number of primary qubits $N_P$ and with higher-depth circuits (i.e. increasing number of resets $N_r$), demonstrating that we approached the asymptotic steady states of these systems in reasonably short order. These results also suggest that this method could prove to be an efficient shortcut to simulating thermal distributions in fault tolerant machines. All that said, it is important to emphasize that even in the noise-free case (where there was no error in the primary qubits), the temperature did not reach zero in our algorithm as formulated, due to stray excitations created by off-resonant interactions between the primary and shadow qubits, and the topological character of excitations in the TFIM specifically.

In future work, it would be interesting to further explore the fine tuning of algorithm parameters with both noise and system size, though the fact that our results converged to fixed approximation ratios and average temperatures suggest that scaling a single parameter set for each model and error rate works well in practice. More illuminating would be the application of this thermometry algorithm in real quantum hardware, and in systems with more complex geometries. We only explored one-dimensional problems in this work to ensure we could get decent scaling estimates from numerical simulations, but our algorithm can be applied in any dimension. On real quantum hardware with long ranged connectivity (either naturally or through error corrected encodings), this method could even be applied to infinite range problems such as quantum spin glasses. In the worst case, the thermalization timescales in such systems are expected to be very long at large $N$, but with rich attendant physics that could be probed in new ways through our methods.

\section*{Acknowledgements}

We would like to thank Erez Berg, Vadim Oganesyan, Eleanor Rieffel, David Schuster, Norm Tubman and Paul Varosy for valuable discussions of the issues in this work. This work was supported by the Superconducting Quantum Materials and Systems Center (SQMS) under contract number DE-AC02-07CH11359. The SQMS Center supported EK's advisory role in this project, and ALS and GG's theoretical and computational research. This work was also supported by the National Science Foundation through grant PHY-1653820, and the Army Research Office through Grant No. W911NF-17-S0001.

\bibliography{references}

\end{document}